\documentclass[aps,prl,twocolumn,superscriptaddress]{revtex4}
\pdfoutput=1
\usepackage{graphicx}
\usepackage{textcomp}
\begin{document}

\title{Nanomechanical motion measured with precision beyond the standard quantum limit}

\author{J. D. Teufel}
\affiliation{JILA, National Institute of Standards and Technology
and the University of Colorado, Boulder, CO 80309, USA}
\author{T. Donner}
\affiliation{JILA, National Institute of Standards and Technology
and the University of Colorado, Boulder, CO 80309, USA}
\author{M.\ A.\ Castellanos-Beltran}
\affiliation{JILA, National Institute of Standards and Technology
and the University of Colorado, Boulder, CO 80309, USA}
\affiliation{Department of Physics, University of Colorado, Boulder,
CO 80309, USA}
\author{J. W. Harlow}
\affiliation{JILA, National Institute of Standards and Technology
and the University of Colorado, Boulder, CO 80309, USA}
\affiliation{Department of Physics, University of Colorado, Boulder,
CO 80309, USA}
\author{K. W. Lehnert} \email{konrad.lehnert@jila.colorado.edu}
\affiliation{JILA, National
Institute of Standards and Technology and the University of
Colorado, Boulder, CO 80309, USA} \affiliation{Department of
Physics, University of Colorado, Boulder, CO 80309, USA}

\date{\today}
\maketitle

\textbf{Nanomechanical oscillators are at the heart of
ultrasensitive detectors of force  \cite{mamin2001}, mass
\cite{jensen2008} and motion
\cite{lahaye2004,anetsberger2009,knobel2003,poggio2008,etaki2008}.
As these detectors progress to even better sensitivity, they will
encounter measurement limits imposed by the laws of quantum
mechanics. For example, if the imprecision \cite{caves1987} of a
measurement of an oscillator's position is pushed below the standard
quantum limit (SQL), quantum mechanics demands that the motion of
the oscillator be perturbed by an amount larger than the SQL.
Minimizing this quantum backaction noise and nonfundamental, or
technical, noise requires an \emph{information} efficient
measurement. Here we integrate a microwave cavity optomechanical
system \cite{regal2008} and a nearly noiseless amplifier
\cite{caves1982,castellanos-beltran2008} into an interferometer to
achieve an imprecision below the SQL. As the microwave
interferometer is naturally operated at cryogenic temperatures, the
thermal motion of the oscillator is minimized, yielding an excellent
force detector with a sensitivity of 0.51\,aN/$\sqrt{\mathrm{Hz}}$.
In addition, the demonstrated efficient measurement is a critical
step towards entangling mechanical oscillators with other quantum
systems \cite{vitali2007}.}

\begin{figure}[!ht]
\includegraphics[width=84mm]{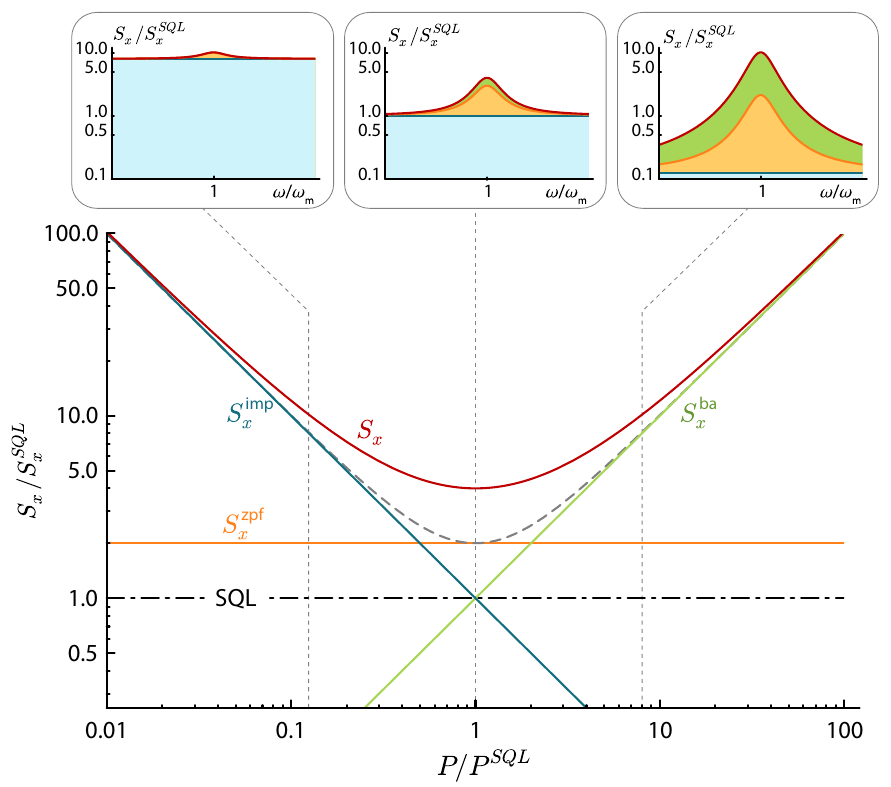}
\caption{\label{fig_introduction} \textbf{Ideal displacement
measurement.} Displacement spectral density $S_{x}$ in units of the
SQL  for temperature $T=0$. $S_{x}$ (red) has three contributions:
the zero-point fluctuations of the beam ($S_{x}^{\mathrm{zpf}}$,
orange), the detector noise ($S_{x}^{\mathrm{imp}}$, blue) given by
the shot-noise limit, and the backaction ($S_{x}^{\mathrm{ba}}$,
green) due to quantum fluctuations of the measurement signal acting
back on the oscillator. At finite temperature there would also be a
thermal contribution. The standard quantum limit (SQL, dashed-dotted
line) is reached at the optimum power $P^{\mathrm{SQL}}$ where
$S_{x}^{\mathrm{imp}}$ and $S_{x}^{\mathrm{ba}}$ contribute equally
such that the noise added by the measurement (grey-dashed line) is
minimal. The main graph shows the contributions to $S_x$ as a
function of power; the subfigures display them as a function of
frequency for three different powers. Here, the white-noise
background is the imprecision or apparent motion while the
Lorentzian peak corresponds to the real motion, comprised of the
zero-point and backaction motions (and the thermal motion for
$T>0$).}
\end{figure}

The quantum mechanical principle that an increasingly precise
measurement of an oscillator's position be accompanied by an
increasingly large force acting back on that oscillator, sets the
SQL as a natural scale for quantifying the noise of a measurement
\cite{caves1980a}. The SQL is the spectral density of the apparent
motion in an ideal measurement, when the backaction motion
$S_{x}^{\mathrm{ba}}$ and the imprecision $S_{x}^{\mathrm{imp}}$ are
equal (see Fig. \ref{fig_introduction}); it corresponds to a quarter
quantum of mechanical noise energy $\hbar \omega_{\mathrm{m}}/4$ and
is given by $S_{x}^{\mathrm{SQL}}=\hbar/\left(m
\omega_{\mathrm{m}}\gamma_{\mathrm{m}}\right)$, where $m$ is the
mass, $\omega_{\mathrm{m}}$ is the mechanical resonance frequency,
and $\gamma_{\mathrm{m}}$ is the mechanical damping rate of the
oscillator.

\begin{figure*}[!ht]
\includegraphics{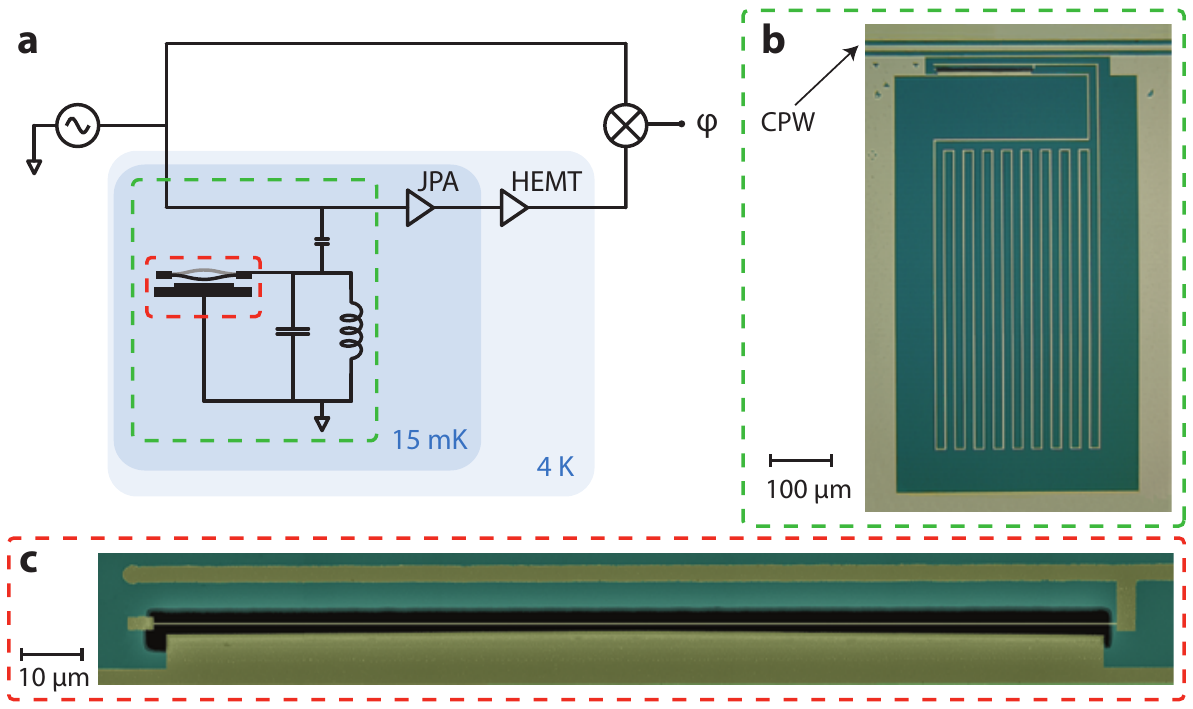}
\caption{\label{fig_setup} \textbf{Interferometric measurement
schematic.} \textbf{a,} The microwave cavity (green-dashed box) is
represented by an LC resonator which is coupled capacitively to the
interferometer. While the inductance is fixed, the total capacitance
depends on the position of a mechanical oscillator (red-dashed box).
The motion of the mechanical oscillator thus modulates the cavity
resonance frequency and a phase shift $\varphi$ is imprinted on a
microwave signal. After amplification with a JPA at 15~mK and a HEMT
amplifier at 4~K, $\varphi$ is measured with a homodyne detection
scheme. \textbf{b,} This false-color optical image shows the
microwave resonant circuit fabricated from aluminum (light brown) on
a silicon substrate (green). The nanomechanical oscillator is
integrated near the coplanar waveguide (CPW) which couples the
interferometer to the cavity optomechanical system. \textbf{c,}
False-color scanning electron micrograph of the aluminum wire freely
suspended over a hole (black) in the substrate.}
\end{figure*}

An interferometer can, in principle, realize the ideal displacement
measurement described in Fig. \ref{fig_introduction} by encoding the
oscillator's displacement into the phase of a light field
\cite{caves1981}. This phase response to motion can be resonantly
enhanced by integrating a cavity containing the mechanical
oscillator into the interferometer \cite{abbott2004,arcizet2006,
corbitt2007a,anetsberger2009,regal2008,groeblacher2009a,thompson2008}.
The imprecision of such a cavity optomechancial measurement is
\begin{equation}\label{Sximp}
    \frac{S_x^{\mathrm{imp}}}{S_x^{\mathrm{SQL}}} =
    \frac{n_{\mathrm{add}}+\frac{1}{2}}{P/(\hbar
    \omega_{\mathrm{c}})} \cdot \frac{1}{\left(\partial \varphi / \partial
    x\right)^2} \cdot \frac{m \omega_{\mathrm{m}} \gamma_{\mathrm{m}}}{\hbar}\,,
\end{equation}
where the first term reflects the phase sensitivity of an
interferometer which adds $n_{\mathrm{add}}$ quanta of noise while
reading out the phase at a rate of $P/(\hbar \omega_{\mathrm{c}})$
photons per second; $P$ is the incident power, and
$\omega_{\mathrm{c}}$ is the cavity's resonance frequency. The
second term describes the transduction of displacement $x$ of the
mechanical oscillator into phase $\varphi$. It is proportional to
$(\gamma_{\mathrm{c}}/g)^2$, where $\gamma_{\mathrm{c}}$ is the
linewidth of the cavity and $g=\partial\omega_{\mathrm{c}}/\partial
x$ is the coupling between the oscillator's motion and the cavity's
resonance frequency. The last term puts the imprecision in units of
$S_x^{\mathrm{SQL}}$. Equation \ref{Sximp} demonstrates that the
absolute imprecision is reduced by measuring with larger power,
stronger coupling, and minimum added noise. The relative imprecision
compared to the SQL benefits from low-mass and high-quality factor
$Q_{\mathrm{m}} =\omega_{\mathrm{m}}/ \gamma_{\mathrm{m}}$
mechanical oscillators.

Our experiments use a microwave cavity optomechanical system and
combine the desirable properties of nanomechanical oscillators with
the simplicity and ideality of an interferometric measurement.
Optical interferometers can operate at the shot-noise limit
($n_{\mathrm{add}}=0$), but are coupled to larger oscillators. While
recent experiments couple nanomechanical objects to evanescent
optical fields \cite{li2008,eichenfield2009,anetsberger2009}, only
interferometers incorporating stiffer or heavier oscillators achieve
the best absolute displacement sensitivity
\cite{arcizet2006,schliesser2008} and have recently approached the
SQL \cite{arcizet2006,corbitt2007a,schliesser2008, thompson2008,
groeblacher2009a}. Our alternative strategy is to use microwave
circuits which naturally couple to nanomechanical oscillators
\cite{regal2008}. This method is especially beneficial for sensitive
force measurements and for observing the quantum behavior of
mechanical oscillators \cite{vitali2007} because it is compatible
with ultralow temperatures. In addition, microwave cavities can be
built as lithographically fabricated integrated circuits and thus
provide a flexible architecture for measuring mechanical motion and
coupling motion to other quantum systems \cite{wallraff2004}. Until
recently, the best option for measuring microwave fields was a
cryogenic high electron mobility transistor (HEMT) amplifier, but
this amplifier contributes more than 20 quanta of noise to
$n_{\mathrm{add}}$, i.e. about 40 times the vacuum noise. This added
microwave noise translates directly into a degraded imprecision for
the mechanical motion \cite{regal2008}. To overcome this added
noise, we developed a nearly shot-noise-limited microwave
interferometer which we use to infer the thermal motion of a
nanomechanical oscillator with an imprecision below the SQL. The key
element of this system is a degenerate Josephson parametric
amplifier (JPA) capable of measuring microwave fields with
$n_{\mathrm{add}}< 1/2$.

\begin{figure*}[!ht]
\includegraphics[width=\textwidth]{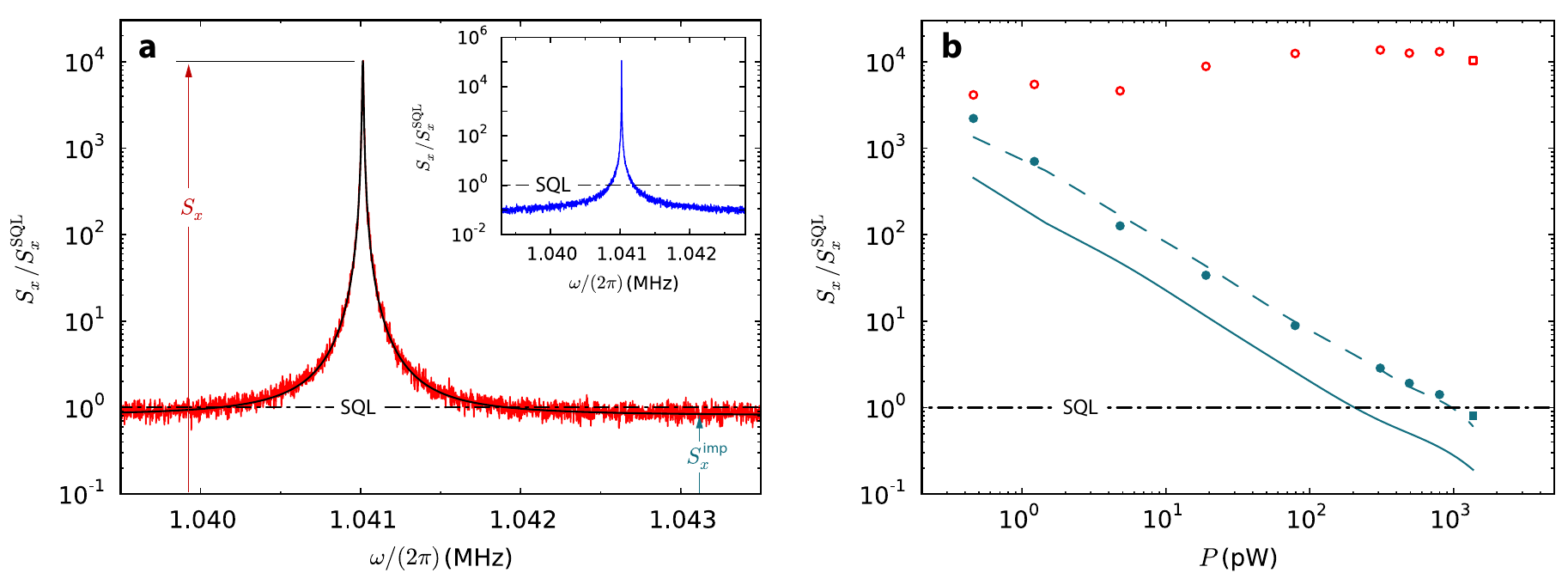}
\caption{\label{fig_Sx} \textbf{Displacement sensitivity below the
standard quantum limit.} \textbf{a,} Shown in red (with a Lorentzian
fit) is the spectral density of the displacement fluctuations
$S_{x}$ in units of the standard quantum limit plotted versus the
Fourier frequency $\omega$. The imprecision (white-noise background)
of this measurement lies below the SQL (black dashed-dotted line).
In the inset, we use passive feedback to increase
$S_{x}^{\mathrm{SQL}}$, improving the imprecision compared to the
SQL, but at the cost of amplified motion. \textbf{b,} Imprecision
$S_{x}^{\mathrm{imp}}$ (blue-filled data points) and total noise
$S_{x}$ (red-open data points) in units of the SQL as a function of
incident microwave power. The data points at the highest power
(shown as squares) were extracted from the data shown in Fig.
\ref{fig_Sx}a. The statistical error is smaller than the size of the
symbols. The blue continuous line shows the theoretical prediction
for $S_{x}^{\mathrm{imp}}$ with a shot-noise-limited interferometer
given our measured mechanical properties. The blue-dashed line is
the theoretical imprecision including the noise added by the entire
measurement chain, independently measured to be $n_{\rm{add}}=1.3$.}
\end{figure*}

We detect the motion of the oscillator with a microwave analog of a
Mach-Zehnder interferometer (Fig. 2). A microwave tone is split and
injected into the two arms of the interferometer. One arm forms the
phase reference; the other arm contains the cavity optomechanical
system which is a microwave resonant circuit with an embedded
nanomechanical oscillator. Variation of the oscillator's position
alters the capacitance of the electrical circuit and therefore its
resonance frequency. The microwave tone excites the circuit and
acquires a phase shift proportional to the mechanical motion. That
phase-modulated microwave signal is first amplified with the JPA and
a cryogenic HEMT amplifier and then mixed with the phase reference,
completing the microwave interferometer. This sensitive phase
measurement allows us to fully characterize the electrical and
mechanical properties \cite{regal2008,teufel2008b} of the cavity
optomechanical system in a dilution refrigerator at temperatures
below 100\,mK.

As in our previous work, we realize a microwave cavity
optomechanical system where the mechanical degree of freedom is the
fundamental mode of a freely suspended aluminum wire. In contrast to
our earlier work \cite{teufel2008a,regal2008,teufel2008b}, we embed
a longer wire into a smaller microwave cavity. As the wire's motion
now varies a larger fraction of the circuit's total capacitance, we
achieve a much larger coupling: $g=2\pi \times 32$~kHz/nm. The wire
has dimensions of 150~\textmu m$\times$170~nm$\times$160~nm,
yielding a total mass $m=(11 \pm 2)$~pg and a mechanical resonance
frequency $\omega_{\mathrm{m}}=2\pi~\times$~1.04~MHz. It is
co-fabricated within an aluminum superconducting inductor-capacitor
(LC) resonant circuit, which behaves as a microwave cavity with
$\omega_{\mathrm{c}}=2\pi\times 7.49$~GHz. The cavity linewidth
$\gamma_{\mathrm{c}}=2\pi~\times$~2.88~MHz is designed so that it is
comparable to $\omega_{\mathrm{m}}$ to facilitate the best possible
displacement sensitivity.

Having determined these system parameters, we can deduce the
displacement fluctuations of the nanomechanical wire from the phase
fluctuations at the output of the interferometer. Figure
\ref{fig_Sx} shows the displacement noise spectral density of the
wire in units of the SQL. Figure \ref{fig_Sx}a displays our lowest
achieved imprecision of $(0.80 \pm 0.03) \times S_x^{\mathrm{SQL}}$,
or 0.2 mechanical quanta. In absolute units, this imprecision is
$S_x=\left[(4.8\pm0.4)\,\mathrm{fm}/\sqrt{\mathrm{Hz}}\right]^2$.
The uncertainty in the absolute imprecision is dominated by the
uncertainty in the mass. However, the determination of the
imprecision compared to the SQL is independent of mass. From the
peak height above $S_x^{\mathrm{imp}}$, we infer the temperature
corresponding to the wire's real thermal motion to be 130~mK, or
2600 mechanical quanta. Note that, in principle, the imprecision can
be arbitrarily small compared to the SQL. For example, the
imprecision compared to the SQL can be artificially reduced by using
feedback to apparently increase $Q_{\mathrm{m}}$. We use passive
feedback from radiation pressure forces \cite{teufel2008b} to
increase $Q_{\mathrm{m}}$ to $1.5\times10^6$ (inset Fig.
\ref{fig_Sx}a), increasing the value of $S_x^{\mathrm{SQL}}$ such
that $S_x^{\mathrm{imp}}=0.07 \times S_x^{\mathrm{SQL}}$. However,
the absolute imprecision remains the same. Furthermore, feedback
amplifies the thermal motion to 2.5\,K. Because this artificial
improvement in imprecision has come at the cost of a proportionally
larger total displacement noise, we now focus solely on the
intrinsic imprecision, i.e. in the absence of feedback.

To demonstrate that the intrinsic imprecision of our interferometer
is close to the shot-noise limit and that we can account for
deviation from this limit, we measure the displacement noise as a
function of incident power. In Fig. 3b we show the power dependence
of both the imprecision $S_x^{\mathrm{imp}}$ (blue-filled symbols)
and of the total displacement noise $S_x$ (red-open symbols)
extracted from noise spectra such as that shown in Fig.
\ref{fig_Sx}a. We also display the predictions for the imprecision
according to equation (\ref{Sximp}). The shot-noise limit is shown
as a solid blue line. From independent measurements, we estimate
that the noise added by our microwave interferometer is reduced from
$n_{\mathrm{add}}=55$ when amplifying with the HEMT only to
$n_{\mathrm{add}}=1.3$ when using the JPA (see Methods and ref 13).
The reduction in noise provided by the JPA is equivalent to an
improvement from 0.9\,\% to 27\,\% in the quantum efficiency of a
photodetector reading out an ideal interferometer. The expected
imprecision for 27\,\% quantum efficiency, or
$n_{\mathrm{add}}=1.3$, is plotted as a dashed line in Fig.
\ref{fig_Sx}b, showing that equation (\ref{Sximp}) accurately
predicts the observed imprecision without adjusting any parameters.
The predictions of equation (\ref{Sximp}) are not straight lines
because they account for the measured values of
$\gamma_{\mathrm{c}}$ and $\gamma_{\mathrm{m}}$ which vary with $P$.
Generally, $\gamma_{\mathrm{m}}$ increases with $P$. This effect and
other power-dependent sources of technical noise (Supplementary
Information) prevent further improvement in the imprecision. They
also account for the larger total displacement fluctuations $S_x$ at
higher power. Because technical noise increases with $P$, a quantum
efficient measurement which minimizes the power needed to reach a
given imprecision is always desirable.

\begin{figure}[!ht]
\includegraphics[width=85mm]{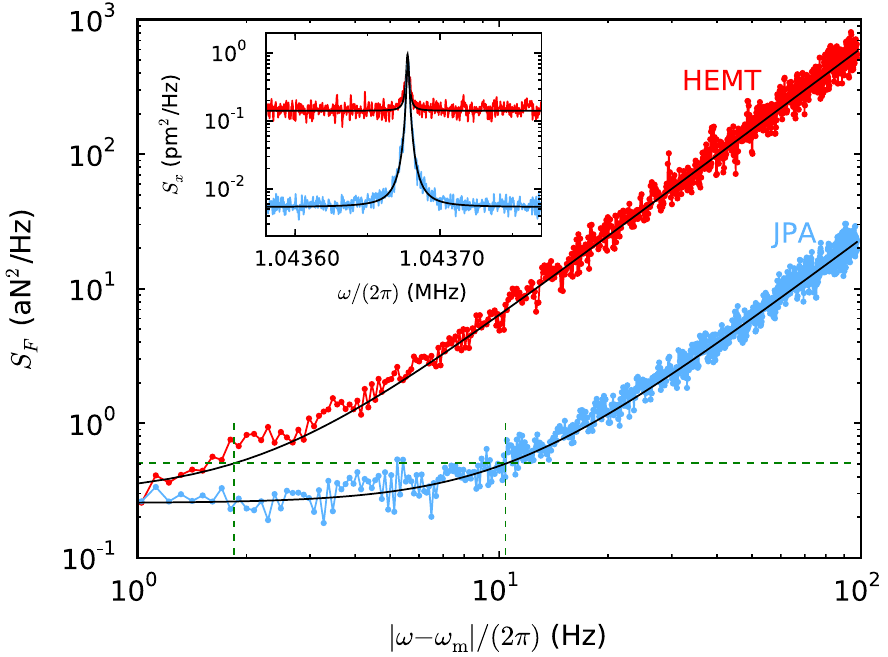}
\caption{\label{fig_forcesensitivity} \textbf{Force sensitivity.}
The force spectral density $S_F$ is shown as a function of
$|\omega-\omega_{\mathrm{m}}|/2\pi$ for a measurement with the HEMT
amplifier only (red) and with the JPA (blue).  The inset shows the
displacement power spectral density from which these force spectra
are inferred, demonstrating the large improvement in displacement
imprecision when using the JPA. As the imprecision is negligible in
comparison to the thermal motion with and without the JPA, the force
sensitivity is not significantly improved. However, there is a
fivefold increase in the bandwidth of the force sensor with the JPA,
corresponding to the larger frequency range over which the
sensitivity is not limited by the imprecision (indicated by the
dashed green lines). At the low microwave power used in this
measurement ($P=1$~pW), $T=77$~mK, $Q_{\mathrm{m}}=6.2\times10^{5}$,
and the resulting force sensitivity on resonance is
$S_{F}=(0.26\pm0.04 )$~aN$^{2}$/Hz, where the dominant uncertainty
comes from that of the mass.}
\end{figure}

In addition to reducing the power required to reach the SQL, the JPA
also reduces the power at which the imprecision becomes less than
the thermal motion of the oscillator. Once the imprecision is small
compared to the thermal motion, a high-$Q$ nanomechanical oscillator
operated at cryogenic temperatures becomes an excellent force sensor
\cite{mamin2001}. The sensitivity to a force applied at the center
of the wire is $S_F=2 S_x/|H(\omega)|^2$, where the mechanical
susceptibility is
$H(\omega)^{-1}=m_{\mathrm{eff}}(\omega^2-\omega_{\mathrm{m}}^2-i
\gamma_{\mathrm{m}}\omega)$, and $m_{\mathrm{eff}}=m/2$ is the
effective mass of the fundamental mode. The force sensitivity on
resonance is $S_{F}=4k_{\mathrm{B}}T
m_{\mathrm{eff}}\gamma_{\mathrm{m}}$, where $T$ is the temperature
corresponding to the thermal motion of the oscillator. Because both
$T$ and $\gamma_{\mathrm{m}}$ are smaller for lower power, the
optimal force sensitivity occurs at a power much less than that
needed to reach the SQL. Figure \ref{fig_forcesensitivity} shows the
displacement and force spectral densities of the oscillator,
measured with and without the JPA.  Measuring with the JPA improves
the displacement imprecision by a factor of more than 28 (inset Fig.
\ref{fig_forcesensitivity}).  As the dominant source of noise near
$\omega_{\mathrm{m}}$ is the thermal motion, we achieve a record
force sensitivity of $S_F=\left[(0.51 \pm
0.04)\,\mathrm{aN}/\sqrt{\mathrm{Hz}}\right]^2$ on resonance, both
with and without the JPA. However, using the JPA increases the
bandwidth of the force sensor by a factor of 5.

In addition to practical applications \cite{degen2009} enabled by
this exquisite force sensitivity, the combination of a highly
efficient microwave interferometer with a high-Q and low-mass
mechanical oscillator opens the route to new experiments probing the
quantum nature of tangible objects. The long-standing goal of
preparing a mechanical object in its ground state implicitly
requires a measurement sensitive enough to resolve the residual
quantum motion. We previously demonstrated the ability of our
microwave cavity optomechanical system to cool the nanomechanical
oscillator using dynamical backaction \cite{teufel2008b} in the
resolved-sideband regime. In this regime, it is, in principle,
possible to cool to the ground state \cite{teufel2008a}. With the
excellent imprecision presented in this work, the zero-point motion
of the oscillator would be easily resolvable above the measurement
background. Once the mechanical oscillator is in its ground state
and sensed with a quantum-limited measurement, a broad field of
challenging and inspiring experiments would be accessible. These
experiments include the generation of squeezed states of the
mechanical mode \cite{woolley2008}, the use of mechanics to produce
squeezed states of the light field \cite{mancini1994}, and the
creation of entanglement between mechanical motion and other quantum
systems \cite{mancini2002,vitali2007}. Even tests of quantum theory
itself have been discussed for optomechanical systems
\cite{marshall2003}.

\appendix*
\section{Methods}
To estimate the contributions to the added noise, we use a
calibrated and variable source of microwave noise power
\cite{castellanos-beltran2008}. We infer that the JPA adds 0.3
microwave quanta of noise and that the HEMT amplifier alone adds
24.5 microwave quanta of noise to our measurement. When operated
with the JPA, the contribution of the HEMT amplifier's added noise
is divided by the JPA's gain, typically 20\,dB. Including the losses
introduced by circulators and transmission lines between the cavity
optomechanical circuit and the different stages of amplification, we
estimate that the interferometer adds 1.16 microwave quanta of noise
to the measurement. The input signal is heavily attenuated by more
than 50\,dB at cryogenic temperatures to reduce its thermal noise to
about 0.17 microwave quanta of noise in addition to the half
microwave quantum of vacuum noise. In total, we estimate the number
of added quanta to be $n_{\mathrm{add}}=1.3$. Our current design was
chosen to allow us to simultaneously measure multiple cavity
optomechanical devices, at the cost of not measuring half of the
microwave power exiting the cavity. This effect is accounted for in
the $\partial \varphi/\partial x$ term of equation (\ref{Sximp}).
However, a simple modification of the geometry would allow us to
measure all of the exiting power. For the same amount of energy
stored in the resonator, this modification would improve the
imprecision by a factor of two.

\begin{acknowledgments}
We acknowledge support from the National Science Foundation's
Physics Frontier Center for Atomic, Molecular and Optical Physics
and from the National Institute of Standards and Technology. T.\ D.\
acknowledges support from the Deutsche Forschungsgemeinschaft (DFG).
We thank N.\ E.\ Flowers-Jacobs for valuable conversations and
technical assistance, and K.\ D.\ Irwin, G.\ C.\ Hilton, and L.\ R.\
Vale for fabrication, and help with the design, of the JPA. K.\ W.\
Lehnert is a member of NIST's Quantum Physics Division.
\end{acknowledgments}

\end{document}